\documentclass{elsart}

\usepackage{graphicx}

\usepackage{amssymb}

\begin{document}

\begin{frontmatter}



\title{An iterative procedure to obtain inverse response functions
       for thick-target correction of measured charged-particle spectra}


\author[inf]{S. Pomp\corauthref{cor1}},
\ead{Stephan.Pomp@tsl.uu.se}
\corauth[cor1]{Corresponding author. Tel.: +46-18-4716850; fax: +46-18-4713853}
\author[inf,cmu]{U. Tippawan}

\address[inf]{Department of Neutron Research, Uppsala University,
              Box 525, SE-751 20 Uppsala, Sweden}

\address[cmu]{Fast Neutron Research Facility, Faculty of Science, Chiang Mai University,
              P.O.Box 217, Chiang Mai 50200, Thailand}

\begin{abstract}
A new method for correcting charged-particle spectra for thick target 
effects is described. 
Starting with a trial function, inverse response functions are found by an 
iterative procedure. 
The variances corresponding to the measured spectrum are treated similiarly
and in parallel.
Oscillations of the solution are avoided by rebinning the data to finer bins 
during a correction iteration and back to the original or wider binning after 
each iteration. 
This thick-target correction method has been used for data obtained with the MEDLEY facility 
at the The Svedberg Laboratory, Uppsala, Sweden, and is here presented in detail and 
demonstrated for two test cases. 
\end{abstract}

\begin{keyword}
Neutron-induced reactions \sep 
Energy-loss correction \sep
Particle-loss correction \sep
Response function

\PACS 34.50.Bw
\end{keyword}
\end{frontmatter}

\section{Introduction}
\label{intro}

Over the past years, several applications involving fast neutrons 
have developed rapidly, which has lead to an increased demand for high-quality
data on neutron-induced production of light charged-particles.
Examples are 
radiation treatment of cancer~\cite{ore98,sch01,lar95},
soft-error effects in computer memories~\cite{single,cha99},
accelerator-driven transmutation of nuclear waste and energy production~\cite{high},
and determination of the response of neutron detectors~\cite{cec79}.

The MEDLEY facility \cite{dan00,tip04}, based at the 
quasi-monoenergetic neutron beam facility of the
The Svedberg Laboratory in Uppsala, Sweden \cite{klu01,pomp04},
has been constructed to measure double-differential cross sections of 
(n,xp), (n,xd), (n,xt), (n,x$^3$He), and (n,x$\alpha$) reactions for a variety of targets
and at neutron energies of up to 175 MeV.
The detector setup consists of eight three-element 
telescopes mounted inside a 100 cm diameter evacuated reaction chamber. 
Each of the telescopes consists of two fully depleted 
$\Delta E$ silicon surface barrier detectors and a CsI(Tl) 
crystal. 
The thickness of the $\Delta E$ detectors
is in the range of 50-60 $\mu$m for the first one ($\Delta E_1$), 
and 400-500 $\mu$m for the second one ($\Delta E_2$). 
The cylindrical CsI(Tl) crystal, 50 mm long and 40 mm in diameter, 
serves as the $E$ detector. 
The $\Delta E - E$ technique is used to identify the light charged particles,
and cutoff energies as low as 2.5 MeV for protons and 4.0 MeV for alpha particles
have been achieved~\cite{tip04,tip06}.
Recently, MEDLEY has been upgraded with thicker and wider CsI(Tl) crystals
to allow measurements even at the highest neutron energies, which are now,
after an upgrade of the neutron-beam facility \cite{pomp04}, 
available with high intensity.

Compared to proton beams, however, the the intensity of neutron beams is still very low.
This fact has to be balanced by the use of thicker targets (typically 100 to 1000~$\mu$m) 
in order to achieve good statistics within a reasonably short beamtime. 
However, the thickness of the targets causes a non-neglible energy-loss 
and even absorption of the produced charged particles. 
This leads to a distortion of the measured spectra which has to be corrected for.
For alpha particles with an energy close to the cutoff, the necessary correction 
of the measured spectra is of the order of a factor of 10 and sizeable even for energies above 20 MeV.

For this correction, an iterative procedure to obtain the inverse response functions of the system
has been developed and implemented in the computer code TCORR.
This code has been used for data obtained with the MEDLEY setup~\cite{tip04,tip06}.
The details of the correction procedure built into TCORR are described in Sect.~\ref{details}.

The problem of the thick-target correction is introduced in more detail in 
Sect.~\ref{problem}, while 
Sect.~\ref{details} describes the specific correction procedure used in TCORR.
Sect.~\ref{disc} presents the application of TCORR to two test cases and a discussion
of the results.
A summary is given in Sect.~\ref{conc}.

\section{The thick-target problem}
\label{problem}

A formal description of the thick-target correction problem and one way to perform 
it is given in Ref.~\cite{johnson}. 
An overview and a discussion of three different correction methods is found in
Ref.~\cite{soderberg}.
In short, the problem can be formulated in the following way. 
Particles produced with a certain energy $E$ inside the target, will be measured 
as having a spectrum of energies $E'$ with $E' \leq E$. 
These distortions, together with other effects like, e.g., detector efficiency,
are usually called response funtions $R(E',E)$, and describe, for a given setup, 
the distribution of the measured energies $E'$ as a function of the true (initial) energy $E$. 
Following Ref.~\cite{soderberg}, the measured spectrum $M(E)$ is then given by
\begin{equation}
\label{eq1}
                  M(E') = \int_{0}^{\infty} R(E',E) T(E) dE,
\end{equation}
where $T(E)$ denotes the true spectrum. 
€
Discretising the problem leads to a matrix equation $M = RT$, i.e.,
\begin{equation}
\label{eq2}
                  M(E'_j) = \sum_{i\geq j} R(E'_j,E_i) T(E_i),
\end{equation}
where $R(E'_j,E_i)$ are the elements of the matrix $R$, which represents the response functions 
and needs to be inverted in order to reconstruct the true spectrum $T$. 

Some authors obtain the response functions by a Monte Carlo codes, 
taking into account 
the target thickness as well as other effects, like the detector response, etc. 
Here, however, the problem of target correction and detector response can easily be separated. 
In the case of MEDLEY, protons with an energy of more than 50 MeV are detected with an 
efficiency of less than 100 \%.
Due to losses caused by nuclear reactions in the CsI detector, the efficiency to detect 100 MeV protons with their 
correct energy is about 90 \%. 
However, due to the employed $\Delta E-E$ technique, those 100 MeV protons that undergo a nuclear reaction 
and, hence, do not deposite the expected amount of energy in the CsI are not recognised 
as protons and thus cut out of the data sample. 
Therefore, the response function for this problem is just one dimensional, i.e., the correction can be done 
by a factor depending only on energy and particle type. 
Hence, the measured spectra $M$ discussed in this paper are already corrected for the detector response, 
which, in our case, is straightforward, unambigious and does not involve an unfolding procedure, 
and we are only concerned with the thick-target correction.

Following Slypen et al. \cite{slypen}, the thick-target correction can be separated into 
an energy-loss and a particle-loss correction. 
But, while Slypen et al. treat the energy-loss correction by an energy shift $\Delta \overline{E}$, 
which is equal to the mean energy loss of particles created in the target and 
escaping with energy $E'$, 
we allow for a distribution of true energies. 
Thus, we write 
\begin{equation}
\label{eq3}
                  T(E_i) = P(E_i) \cdot \sum_{j=0}^{j=i} R_{inv}(E_i,E'_j) M(E'_j),
\end{equation}
where the matrix $R_{inv}$ describes the distribution of possible true energies $E_i$ 
for a given measured energy $E'_j$. 
The factor $P(E_i)$ stands for the particle-loss correction, 
taking into account that a certain fraction of 
the produced particles are stopped already inside the target. 
Thus, our problem is to find the matrix $R_{inv}$.

Throughout the reminder of this paper, we will loosely denote $R_{inv}$ the inverse response matrix, 
while the $R_{inv}(E_i,E'_j)$ for a fixed $E'_j$ will be called inverse response functions. 

Starting from a first guess, 
improved inverse response matrices are obtained from the true spectrum by an iterative procedure, 
while the $P(E_i)$ remain fixed for a given target material, target thickness and particle type. 

\section{The target correction procedure used in TCORR}
\label{details}

\begin{figure}
 \begin{center}
  \includegraphics*[width=10cm]{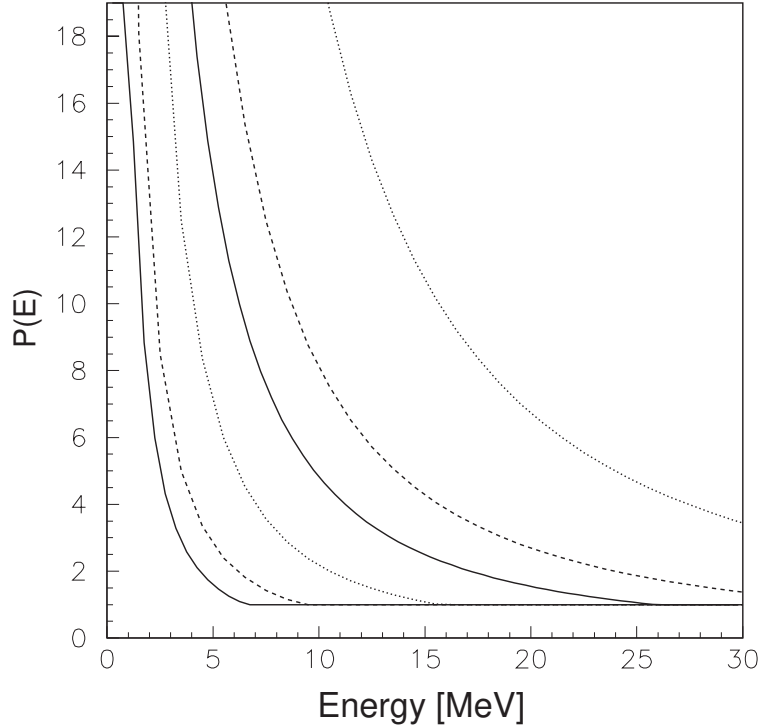}
 \end{center}
  \caption{
    Examples for the particle-loss correction $P(E)$ for protons (left three lines)
    and $\alpha$ particles (right three lines) emitted from a 334 $\mu$m thick silicon target (solid line),
    as well as a 221 $\mu$m thick (dashed line) and a 552 $\mu$m thick (dotted line) iron target.
    The cutoff energy has been set to zero in all cases.
   }
 \label{fig:fig1}
\end{figure}

We start by calculating the particle-loss correction $P(E_i)$
and construct the first guess for the inverse response matrix $R_{inv}$.
With $S(e)$ being the stopping power for the considered particle type in the target, 
the range $X(E)$ of a particle with energy $E$ is given by
\[
      X(E) = \int_{0}^{E} \frac{de}{S(e)}.
\]
If $X(E)$ is larger than the target thickness $t$ along the line-of-sight of the detector, 
no particles are lost inside the target and no correction is needed. 
If, however, $X(E)$ is smaller than $t$, only a fraction $\frac{X(E)}{t}$ of the target allows 
particles with the initial energy $E$ to reach the detector. 
Since the particle identification introduces a cutoff, which is in the order of a few MeV,
only particles emitted from the target with an energy above the cutoff are registered.
Thus, in order to normalise to the same number of target nuclei, 
we obtain for each initial energy $E_i$: 
\begin{equation}
\label{eq4}
      P(E_i) = \left\{ \begin{array}{cll} \frac{t}{X(E_i)-X(E_{cutoff})} \; & \; & \; \mbox{if $X(E_i) < t $}     \\
                                                            1.0          \; &    & \; \mbox{otherwise}
                       \end{array} 
               \right. .
\end{equation}
Figure~\ref{fig:fig1} shows some examples for the particle-loss corrections.

%
%

The first guess for the distribution of true energies $E_i$, i.e., the inverse response function, for a 
measured energy $E'_j$ is,
except for binning effects, a square distribution.
Taking care also of binning effects, we obtain the complete matrix from
the following equation: 
\begin{equation}
\label{eq5}
       R_{inv}^{(0)}(E_i,E'_j) = \left\{ \begin{array}{lcl} \;    0        \; & \; \mbox{for} \; & \; E_i^+ < E'_j \;\; \mbox{or} \;\; E_i^+ > E'_j + \Delta E \\
                                                            \; \frac{1}{k} \cdot \frac{\left| E_i - E'_j \right|}{\Delta E_{bin}}   \; & \mbox{for} & \; E_i^- < E'_j < E_i^+ \;\; \mbox{or} \\ 
                                                                                                                                       &            & \; E_i^- < E'_j + \Delta E_{max}< E_i^+ \\
                                                            \; \frac{1}{k} \; & \mbox{otherwise}  &  
                                          \end{array} 
                                 \right. ,
\end{equation}
with
\begin{eqnarray*}
  \;\;\;\;      k            & \;  = \;  &    \frac{\Delta E}{E_{i+1} - E_i}       \\
      E_i^+        &     =     &    E_i + \frac{1}{2}\Delta E_{bin}      \\
      E_i^-        &     =     &    E_i - \frac{1}{2}\Delta E_{bin}      \\
   \Delta E_{bin}  &     =     &    E_{i+1} - E_i                     .
\end{eqnarray*}
The constant $k$ is the number of bins that are not equal to zero,
$\Delta E_{max}$ is the maximum energy loss for a particle emitted 
from the target with the detected energy $E'_j$,
and $\Delta E_{bin}$ the chosen bin width.
With $S(e)$ being the stopping power for the considered particle type in the target, $\Delta E$ is defined by
\[
      t = \int_{E'_j}^{E'_j + \Delta E} \frac{de}{S(e)}.
\]

\begin{figure}
 \begin{center}
  \includegraphics*[width=10cm]{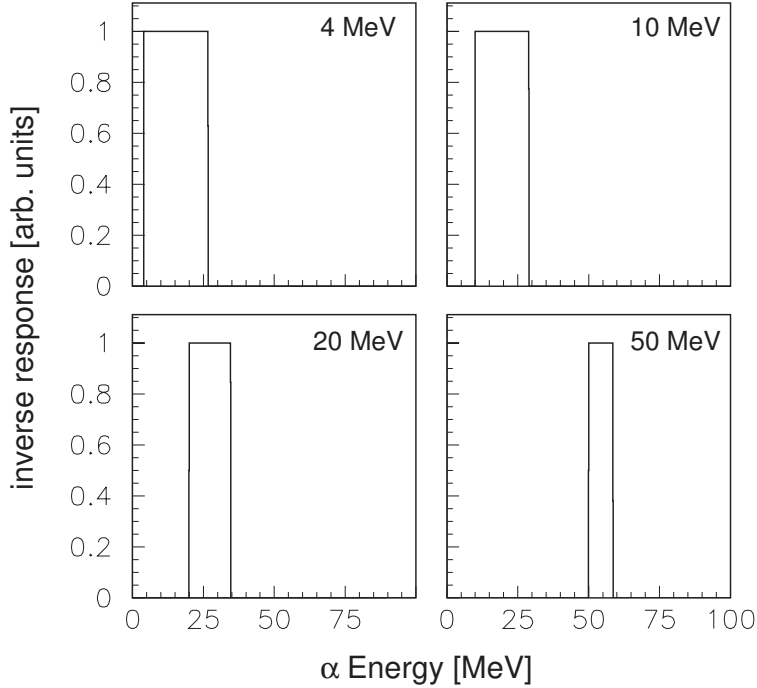}
 \end{center}
  \caption{
    Inverse response functions for $\alpha$ particles from a 334 $\mu$m thick silicon target
    for the measured energies $E'_j$ given in the graphs. 
    The thickness corresponds to the 303 $\mu$m target used in Ref.~\cite{tip04} and an
    emission angle of the particles of 65 degrees relative to the surface plane.
   }
 \label{fig:fig2}
\end{figure}

By construction, the sum of the elements of $R_{inv}$ is, for any given $E'_j$, 
normalised to unity:
\[
\sum_{i} R_{inv}^{(0)}(E_i,E'_j) = 1.
\] 
It has been verified by Monte Carlo simulations that Eq.~\ref{eq5} gives a 
very precise description of the inverse response function since 
no deviations from the assumed square distribution could be found. 
Figure~\ref{fig:fig2} shows four inverse response functions for alpha particles 
originating from a silicon target.

%
%

In Eq.~\ref{eq5} an equal probability for all initial energies $E_i$ is assumed. 
Now, one can start an iteration process by first obtaining a guess for the true spectrum,
based on the current best guess of $R_{inv}$ and using Eq.~\ref{eq3}.
The true spectrum after $n$ iterations is then given by
\begin{equation}
\label{eq7}
                  T^{(n)}(E_i) = P(E_i) \cdot \sum_{j=0}^{j=i} R_{inv}^{(n)}(E_i,E'_j) M(E'_j),
\end{equation}
For the following iterations, improved guesses of $R_{inv}^{(n+1)}$ are obtained from $T^{(n)}$:
\begin{equation}
\label{eq6}
                   R_{inv}^{(n+1)}(E_i,E'_j)  =   c \cdot R_{inv}^{(0)}(E_i,E'_j) \cdot T^{(n)}(E_i).
\end{equation}
The constant $c$ is chosen to ensure normalisation, such that
$
           \sum_{i} R_{inv}^{(n+1)}(E_i,E'_j) = 1
$
for all $E'_j$.
Going back to Eq.~\ref{eq7}, the iteration is continued until a satisfactory solution is found.

In cases where the inverse response functions are wide, i.e., particles with large energy losses,
it is in general favourable to start the iteration procedure not with the original square distributions
$R_{inv}^{(0)}$ but with $R_{inv}^{(1)}$ as obtained from Eq.~\ref{eq6} assuming $T^{(n)}(E_i) = M(E_i)$.
 
The variances corresponding to the measured spectrum are to be treated in almost exactly 
the same way as described above.
The only difference is that the particle-loss correction in Eq.~\ref{eq7} has to enter in quadrature.

The described iteration procedure has been implemented in the FORTRAN code TCORR.
Using the variance histogram, TCORR checks after each iteration, by means of a Kolmogorov test~\cite{kolmo}, 
whether the last change in the true spectrum is so small that the iteration can
be consider as converged and be stopped.

To avoid some of the problems introduced by the, in general, wide binning of the experimental data, 
the spectra are rebinned to much finer binning during each iteration and 
rebinned back to the original bin width after each iteration. 
By this smoothing method, the problem of running into oscillations is also diminished.

\section{Application and discussion of the method}
\label{disc}

The strength of the method is its simplicity due to the use of inverse response functions,
which are, starting from a first guess, improved by an iterative procedure.
The principle idea is similar to the one of Ref.~\cite{slypen},
but while Slypen et al. treat the energy-loss correction in each bin by one energy shift $\Delta \overline{E}$, 
which is equal to the mean energy loss of particles created in the target and escaping 
with energy $E'$, we here allow for a distribution of true energies. 
This is an improvement since $\Delta \overline{E}$ goes towards zero when approaching the cutoff, 
while the particle-loss correction goes to infinity.
Thus, the correction of Slypen et al. tends to overestimate the correction at low energies.
%
%
%
%
The more realistic approach of allowing for a distribution of initial energies resulting in
the same measured energy bin avoids this problem since the width of the distribution
becomes wider when approaching cutoff.

%
%

We have tested the described procedure for many different cases~\cite{uddephd}. 
For experiments concerned with the measurment of light-ion production, 
alpha particles are affected the most by the target.
Therefore, we discuss below two test cases with alpha particle spectra.
The binning of the data entering the target correction procedure in the following examples varies
between 0.5 and 5 MeV.
The inverse response matrix $R_{inv}$, however, is in all cases calculated for energy steps of 0.1 MeV.
Therefore, during each iteration, the spectrum to be corrected is rebinned into 0.1 MeV bins with
equal contents.
The result is then again binned like the experimental input data.

\begin{figure}
 \begin{center}
  \includegraphics*[width=10cm]{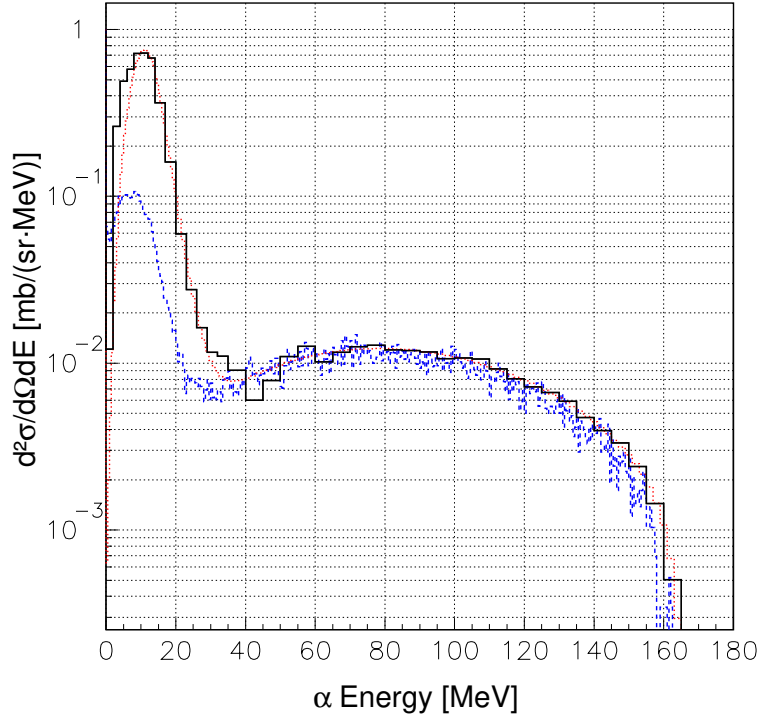}
 \end{center}
  \caption{
    The dotted line shows the energy distribution $T(E)$ for alpha particles 
    as calculated with the TALYS code for the Fe(n,x$\alpha$) reaction with 175 MeV neutrons.
    From this, the $M(E')$ spectrum, assuming a 200 $\mu$m thick target, was obtained
    with TARGSIM (dashed line).
    The detection threshold was set to 0 MeV.
    The solid line shows the result for $T^{(10)}(E)$ as given by TCORR.
   }
 \label{fig:fig3}
\end{figure}

\begin{figure}
 \begin{center}
  \includegraphics*[width=10cm]{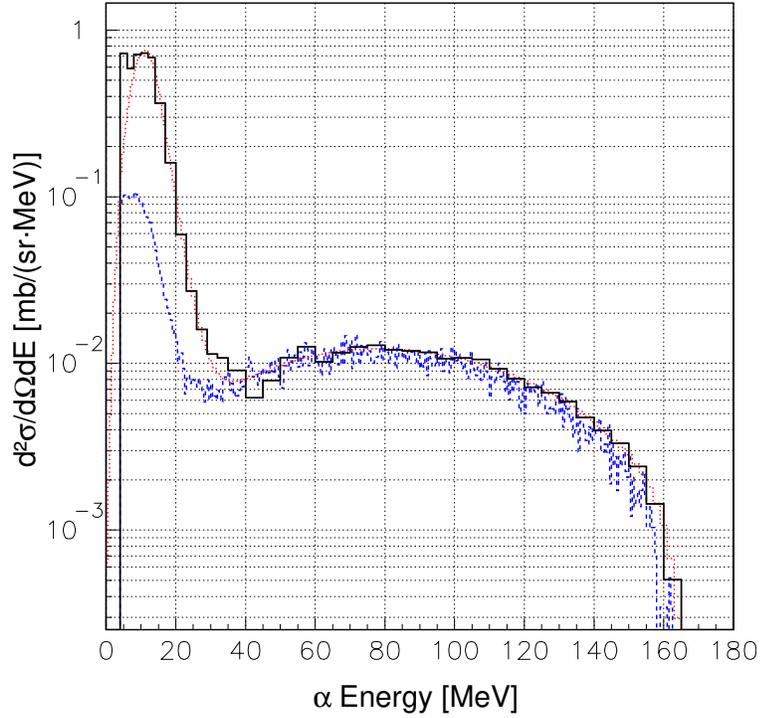}
 \end{center}
  \caption{
    Same as Fig.~\ref{fig:fig3}, but for a detection threshold of 4 MeV.
   }
 \label{fig:fig4}
\end{figure}

The first test case starts with an assumed but realistic true spectrum $T(E)$ for Fe(n,x$\alpha$) at 20 degrees
for an incoming neutron energy of 175 MeV. 
The expected true energy distribution $T(E)$ has been calculated with the TALYS code~\cite{talys} and
is shown as a dotted line in Figs.~\ref{fig:fig3} and~\ref{fig:fig4}.
To simulate the effect of the iron target on the energy spectrum of the alpha particles,
the Monte Carlo code TARGSIM, based on GEANT 3.21~\cite{geant} of the CERN software library, was used.
The TARGSIM code simulated a measured alpha particle energy spectrum $M(E')$, as emitted from a 200 $\mu$m iron target,
tilted by 45 degrees relative to the beam axis, towards the 20 degree telecope.
The effective target thickness is therefore 221 $\mu$m.
In the simulation, 100,000 alpha particle events were produced. 
Out of these, 28,422 alpha particles, or about 28.4 \%, escaped the target.
The remaining 71.6 \% of the alpha particles stopped already inside the target.
The resulting energy distribution is shown as a dashed line in Fig.~\ref{fig:fig3} and 
is the assumed measurement entering the target correction.
The dashed line in Fig.~\ref{fig:fig4} shows the same simulated measured spectrum $M(E')$
but with a detection threshold of 4 MeV.
Finally, the solid line shows the result for $T^{(10)}(E)$ of TCORR, i.e., the reconstructed true spectrum
after ten iterations.
Note the shift in the evaporation peak from about 11 MeV in the TALYS calculation for $T(E)$ down to about 7 MeV
in $M(E')$ due to the target effects, which is corrected back into its right position in $T^{(10)}(E)$.

%
%
%
%
%

\begin{figure}
 \begin{center}
  \includegraphics*[width=10cm]{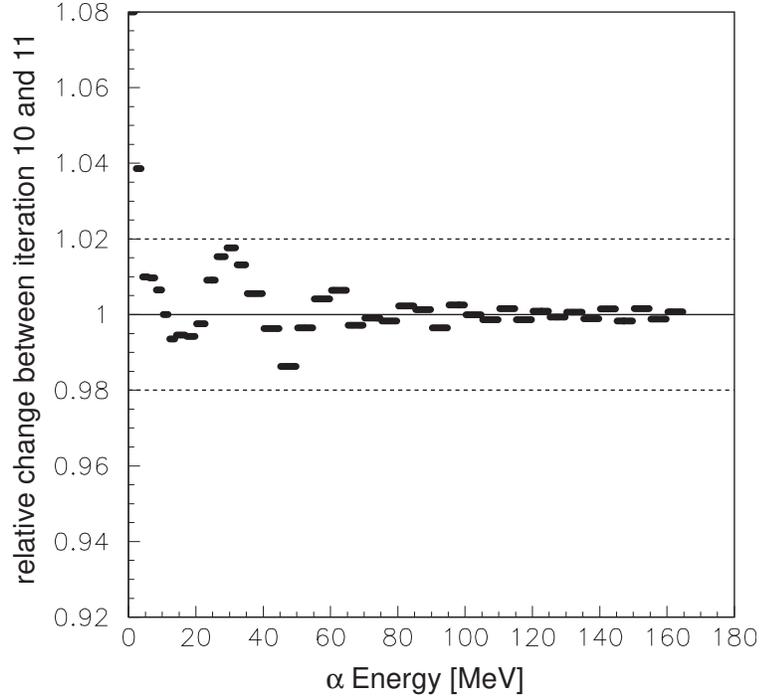}
 \end{center}
  \caption{
    Relative differences between the results $T^{(10)}(E)$ and $T^{(11)}(E)$ of TCORR for
    the 0 MeV detection threshold.
    Note that the relative changes between 0 and 2 MeV are as high as 1.35 and 
    outside the plotted range.
   }
 \label{fig:fig5}
\end{figure}

\begin{figure}
 \begin{center}
  \includegraphics*[width=10cm]{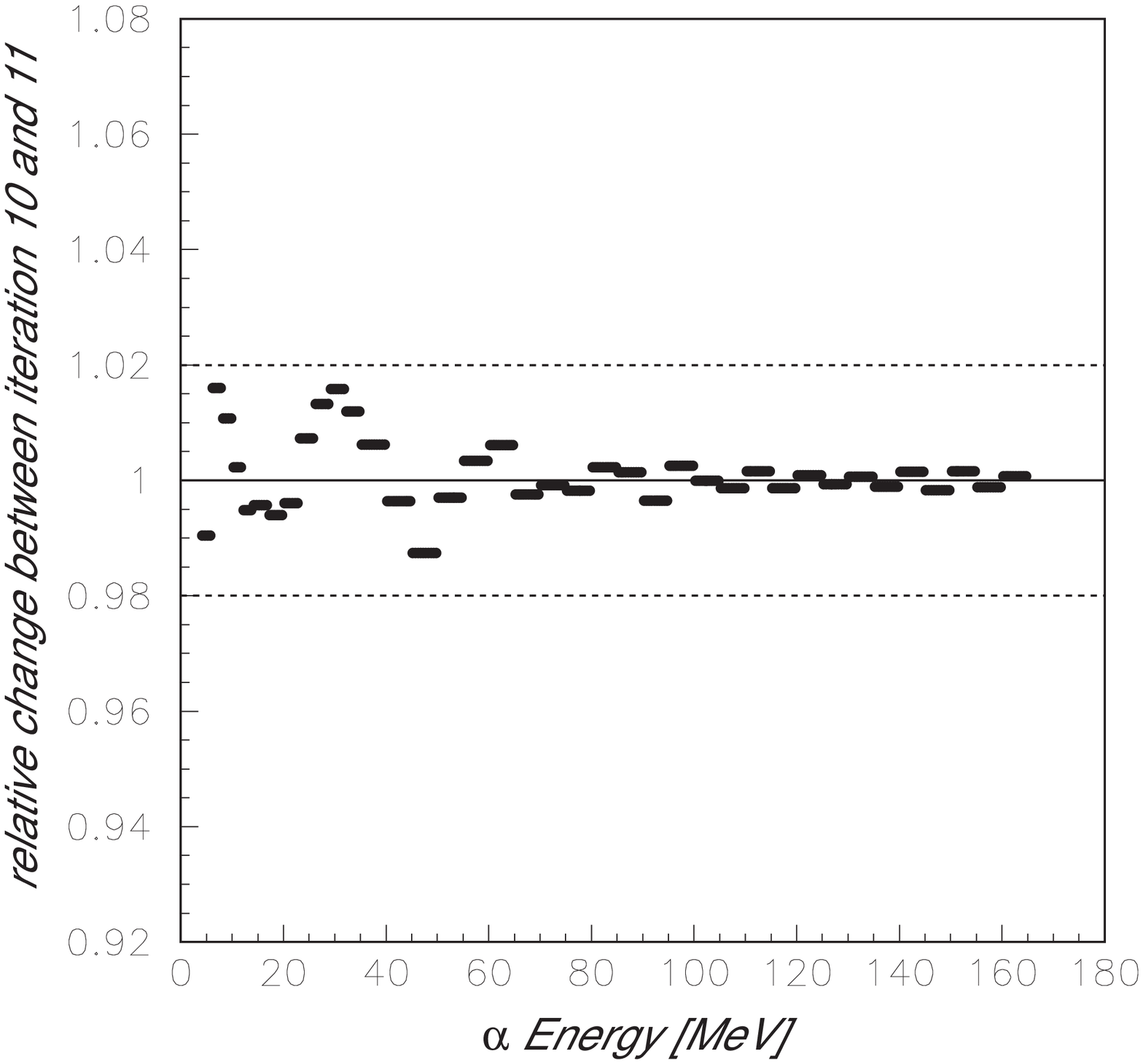}
 \end{center}
  \caption{
    Relative differences between the results $T^{(10)}(E)$ and $T^{(11)}(E)$ of TCORR for
    the 4 MeV detection threshold.
   }
 \label{fig:fig6}
\end{figure}

In this first test case we have not assumed any errors in the measured data.
We therefore just stop the iteration procedure at a point when changes from one iteration to the
next become small as compared to typical experimental errors. 
The relative changes from $T^{(10)}(E)$ to the result after the next iteration, $T^{(11)}(E)$, are shown in 
Figs.~\ref{fig:fig5} and~\ref{fig:fig6} for the two detection thresholds.

The agreement between the original $T(E)$ and $T^{(10)}(E)$ reconstructed by TCORR is good over almost 
the complete energy region but can be off by up to a factor of 5 for some bins next to the assumed detection threshold. 
Here, the particle correction goes towards infinity and the correction procedure becomes very sensible 
to the used input data and to the used energy-range tables.
%

%
%
%
%
%
%

\begin{figure}
 \begin{center}
  \includegraphics*[width=10cm]{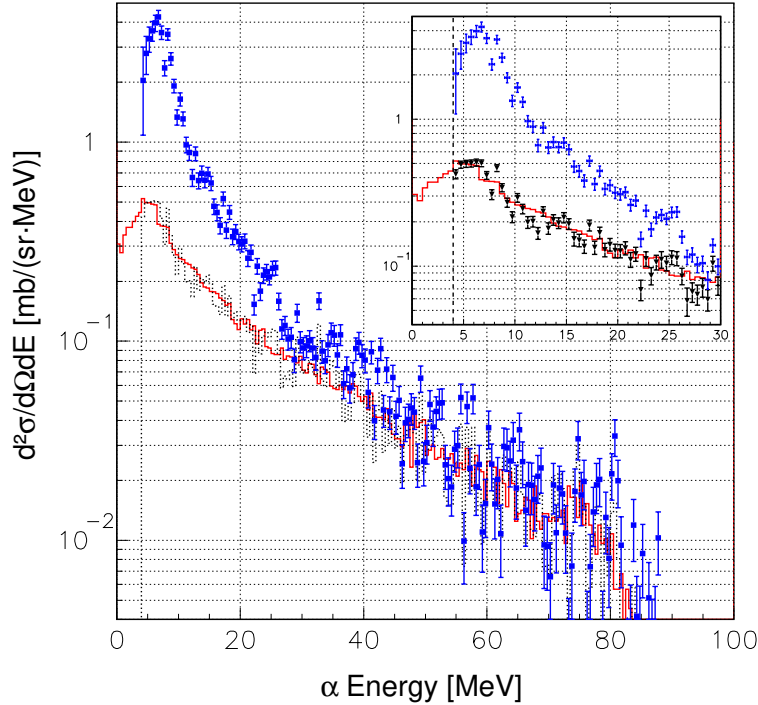}
 \end{center}
  \caption{
    TCORR results for alpha particle spectra using data from Ref.~\cite{tip04}.
    The dotted lines show the measured energy distribution $M(E')$ of the alpha particles in 0.5 MeV bins.
    The square symbols show the TCORR result $T^{(4)}(E)$ with 0.5 MeV bins (left)
    and with bins of varying size of 2 to 5 MeV (right).
    TARGSIM has been used to simulate a measured spectrum $M(E')$ (solid line). 
    The insets show the energy region close to the detection threshold, indicated by the dashed vertical line.
    The true experimental data are this time shown as triangles together with their statistical errors.
%
%
   }
 \label{fig:fig7}
\end{figure}

\begin{figure}
 \begin{center}
  \includegraphics*[width=10cm]{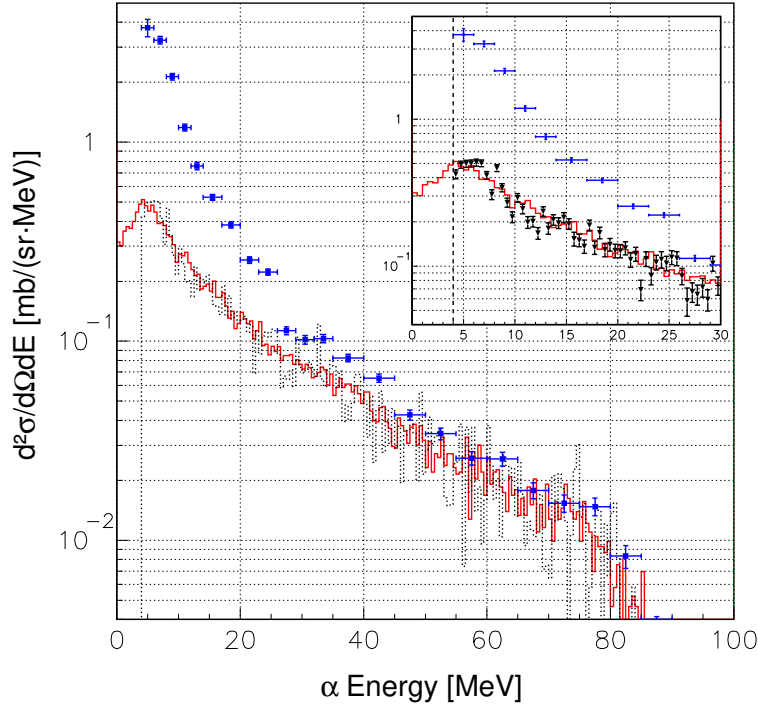}
 \end{center}
  \caption{
    Same as Fig.~\ref{fig:fig7}, but with a bin size varying between 2 and 5 MeV for $T^{(4)}(E)$.
%
%
   }
 \label{fig:fig8}
\end{figure}

For the second case we use experimental data obtained with the MEDLEY setup
for the Si(n,x$\alpha$) reaction with an incoming neutron energy of 96 MeV~\cite{tip04}.
The dotted lines in Figs.~\ref{fig:fig7} and~\ref{fig:fig8} show the measured energy distribution $M(E')$
of alpha particles emitted under 20 degrees and with a detection threshold of 4 MeV and in 0.5 MeV bins.
The used target has an effective thickness of 334~$\mu$m along the line of sight.
The insets show the energy region between 0 and 30 MeV and the experimental data (triangle symbols) together
with their statistical errors. 
The detection threshold is indicated by the dashed vertical lines.

The result of the target correction after four iterations, $T^{(4)}(E)$, is shown by the square symbols 
together with the propagated statistical errors.
In Fig.~\ref{fig:fig7} the target correction was performed without rebinning 
of the experimental data, while in Fig.~\ref{fig:fig7} the data where rebinned in 2 to 5 MeV bins 
to achieve better statistics in each bin.
However, as mentioned above, the inverse response functions were in both cases calculated in steps and bins of 0.1 MeV.
The correction iteration is stopped once the Kolmogorov test gives a probability of more than 99\% that the current 
result and the result from the previous iteration represent the same energy distribution.
In both the shown cases, this was the case for iteration number four.

The solid lines in Figs.~\ref{fig:fig7} and~\ref{fig:fig8} indicate the result of TARGSIM taking 
the target correction results, $T^{(4)}(E)$, as input.
In this simulation of the target effects, also the resulting alpha energy spectrum below 4 MeV is shown.
The simulated and the experimental $M(E')$ agree over the whole energy range in both cases.
%

%
%

%
%
%
%
%
%
%
%

\section{Summary}
\label{conc}

A method for correcting energy spectra distorted by energy and particle loss in a thick target has been described.
It uses an iterative procedure to obtain improved guesses on the inverse response functions for
each observed particle energy.
The method has been implemented into a Fortran code, TCORR, which is used in the analysis 
of several data sets obtained with MEDLEY~\cite{tip04,tip06}.
The procedure is easy to use, includes a correct treatment of cutoff energies,
and has been validated by some test cases.
Variances corresponding to the measured spectra are treated in parallel
and in a similar way.

We have shown two examples for target correction of alpha particle spectra.
In the first case, we have considered alpha particles being emitted from a 221 $\mu$m iron target.
A true spectrum, $T(E)$, was calculated with the nuclear model code TALYS and target effects have been simulated
with the Monte Carlo code TARGSIM.
Correcting the resulting $M(E')$ spectrum with TCORR lead back to a result which is in good agreement 
with the original input data.
In the second case we started with experimental data from MEDLEY for alpha particles emitted from
a 334 $\mu$m silicon target and obtained a true spectrum with TCORR.
Taking these corrected data as input to the target simulation code TARGSIM lead back to an alpha energy spectrum
which is in good agreement with the input data. 
We take this as confirmation that the procedure works and, within the experimental errors,
gives correct results for the true spectrum $T(E)$.

\ack
  We thank
  Profs. Leif Nilsson and Jan Blomgren for their support and valuable comments
  on the manuscript.
  We would also like to thank
  Prof. Valentin Corcalciuc for many discussions on the subject.
This work was supported by 
the Swedish Natural Science Research Council, 
the Swedish Nuclear Fuel and Waste Management Company, 
the Swedish Nuclear Power Inspectorate, 
Ringhals AB, 
the Swedish Defence Research Agency, 
and the Swedish International Development Authority. 






\end{document}